\documentclass[10pt,a4paper]{book}
\usepackage{frontier07}
\begin{document}
\newcounter{ctr}
\setcounter{ctr}{\thepage}
\addtocounter{ctr}{8}

\talktitle{Cosmological WIMPs, Higgs Dark Matter and GLAST}
\talkauthors{A. Sellerholm \structure{a}, 
             J. Conrad \structure{a,b},
             L. Bergstr\"om \structure{a}
             J. Edsj\"o \structure{a} \\
             Representing the GLAST-LAT collaboration}
\begin{center}
\authorstucture[a]{Department of Physics, Stockholm University, Stockholm, Sweden}

\authorstucture[b]{Department of Physics, Royal Institute of Technology (KTH), Stockholm, Sweden}
\end{center}

\shorttitle{EGBR and Dark Matter} 

\firstauthor{A. Sellerholm, J. Conrad, L. Bergst\"rom, J. Edsj\"o}

\begin{abstract}
Measurement of the extragalactic background (EGBR) of diffuse gamma-rays is perhaps one of  the most challenging tasks for future gamma-ray observatories, such as GLAST. This is because any determination will depend on accurate subtraction of the galactic diffuse and celestial foregrounds, as well as point sources. However, the EGBR is likely to contain very rich information about the high energy-gamma ray sources of the Universe at cosmological distances.

We focus on the ability of GLAST to detect a signal from dark matter in the EGBR. We present sensitivities for generic thermal WIMPs and the Inert Higgs Doublet Model. Also we discuss the various aspects of astrophysics and particle physics that determines the shape and strength of the signal, such as dark matter halo properties and different dark matter candidates. Other possible sources to the EGBR are also discussed, such as unresolved AGNs, and viewed as backgrounds.
\end{abstract}

\section{Introduction}
The nature of dark matter (DM) is still one of the most challenging mysteries in present day cosmology and is so far completely unknown. 
The upcoming Gamma Ray Large Area Space Telescope (GLAST) \cite{glast} will survey a previous unexplored window to the high energy $\gamma$-ray universe, playing a crucial role in the indirect detection of weakly interactive massive particle (WIMP) DM through their self annihilation products, resulting in photons. GLAST will pursue different searches for DM, including point source surveys, such as the galactic center, and diffuse emission studies \cite{Conrad} . 

In this paper we focus on the diffuse signal from cosmological, extragalactic WIMPs and prospects for GLAST to detect such a signal. We do this for a typical, thermal WIMP and a specific particle physics scenario with an extended Higgs sector.

Also we examine a recent claim that the EGBR is compatible with a 60 GeV cosmological WIMP \cite{deboer} but where we suspect that the effect of cosmology has not been taken into account in the calculations.    

Any sensitivity calculation is dependent on the background of the signal and we present an estimate of possible astrophysical contributions to the EGBR.  

\subsection{DM candidates}
There exists many extensions of the standard model of particle physics that contain suitable WIMP DM candidates. Usually these are neutral, stable particles with masses and interaction strengths that give the observed, present day relic abundance. Probably the most studied of such particles is the neutralino, the lightest neutral particle that arises in supersymmetric extensions of the standard model (see, e.g., \cite{kameonkowski}) and is often used as the archetype for fermionic DM . The mass range of the neutralino is usually from around 50 GeV to a few TeV. 

The lightest Kaluza Klein excitation (often the first excitation of the hyper charge gauge boson) gives an archetype for  vector bosonic DM with mass in the range of about 0.5 TeV to a few TeV, see for example \cite{hooper} and references therein. Below we shall also discuss an archetype for scalar DM.

\begin{figure}  
\begin{center}
\epsfig{figure=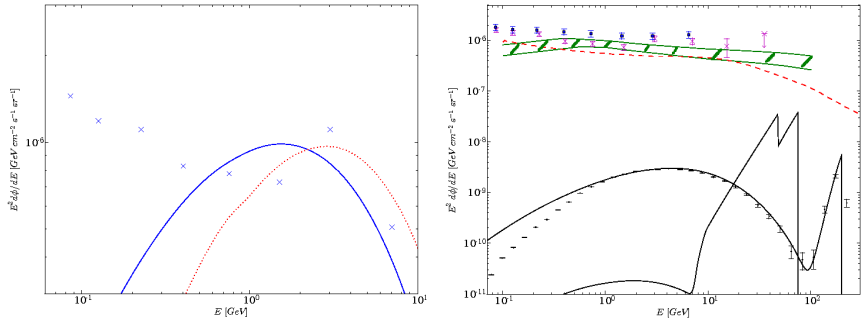,width=11.5cm} 
\end{center}
\caption{Left panel: Comparing the photon spectra at emission (red dotted) to the cosmological signal (blue solid), from a 60 GeV WIMP, see text. Right panel: EGRET data points  (squares from \cite{sreekumar}, crosses from \cite{strong}). The green hatched area represents the upper and lower limits of astrophysical sources contributing to the EGBR, as taken from \cite{dermer}. The red, dashed line is the unresolved blazar model used in our sensitivity calculation. Also we are showing two examples of cosmological WIMP spectra; a 75 GeV IDM WIMP and a 200 GeV WIMP of the kind used for the GLAST sensitivities, the latter also shows the response of GLAST. } 
   \label{deboer}
\end{figure}

 
\paragraph{The Inert Doublet Model} (IDM) \cite{barbieri} is a minimal extension of the
standard model -- an added second Higgs doublet $H_2$, with an
imposed unbroken discrete $Z_2$ symmetry that forbids its direct coupling to
fermions (i.e.\ $H_2$ is \emph{inert}). 
In the IDM the mass of the particle that plays the role of the standard model Higgs can be as high as about 500
GeV and still fulfill present experimental precision tests.
Furthermore, conservation of the $Z_2$ parity implies that the
lightest inert Higgs particle ($H^0$) is stable and hence a good
DM candidate. One of the interesting features of the IDM is that it offers very high annihilation  branching ratios into $\gamma\gamma$ and $Z\gamma$ final states, compared to the branching ratios into quarks, yielding the continuum spectra \cite{gustafsson}. The range of WIMP masses is just in the range where GLAST will be sensitive.  A spectrum from a cosmological IDM WIMP can be seen in figure~\ref{deboer}.

%

%

\section{The cosmological WIMP signal}
The diffuse photon-signal originating from DM annihilating throughout the Universe can be calculated in several ways. Here we follow the procedure of \cite{ullio}, where the number of photons per unit effective area, time and solid angle in the redshifted energy range $E_0$ to $E_0+dE_0$, is given by:
\begin{equation}
\frac{d\phi}{dE_0}=\frac{\sigma v}{8\pi}\frac{c}{H_0}\frac{\bar{\rho}^2_0}{M^2_\chi}\int\,dz\,(1+z)^3\frac{\Delta^2(z)}{h(z)}\frac{dN_\gamma(E_0(1+z))}{dE}e^{-\tau(z,E_0)}.
\label{flux}
\end{equation}
In the following we will discuss the various quantities contributing to eq.~(\ref{flux}).

%
%

\subsection{High energy $\gamma$-ray environment}

Any extragalactic $\gamma$-ray signal is strongly  affected by absorption in the inter-galactic medium, especially at high energies. The absorption is parameterized by the parameter $\tau$, the optical depth. The dominant contribution to the absorption in the GeV-TeV energy range is pair production on the extragalactic background light emitted in the optical and infrared range. For the optical depth, as function of both redshift and observed energy, we use the results of \cite{primack}. 
%
Newer calculations of optical depth are now available, see \cite{stecker}. These 
results imply a slightly lower optical depth at low redshifts and slightly higher at high redshifts, which in turn slightly enhances or suppresses the WIMP signal, respectively.


\subsection{Particle physics}

The preferred particle physics model enters the differential gamma-ray flux via the cross section, $\sigma$, the WIMP mass $M_\chi$ and the differential gamma-ray yield per annihilation $dN/dE$, which is of the form:
\begin{equation}
\frac{dN_\gamma}{dE}=\frac{dN_\textrm{cont}}{dE}(E)+2b_{\gamma\gamma}\delta(E-M_\chi)+b_{Z\gamma}\delta(E-M_Z^2/4M_\chi))
\label{yield}
\end{equation}
The first term in eq.~(\ref{yield}) is the contribution from WIMP annihilation into the full set of tree-level final states, containing fermions gauge or Higgs bosons, whose fragmentation/decay chain generates photons. These processes give rise to a continuous energy spectrum.
The second and third terms correspond to direct annihilation into final states of  two photons and of one photon and one  Z boson, respectively. Although of second order (one loop processes), these terms can give rise to significant amounts of monochromatic photons. 

Since the emission spectrum of the continuum and line signal are very different in shape, the result of the integration over redshift, in eq.~(\ref{flux}), is quite different. The continuum spectra becomes slightly broadened and the peak is red shifted to lower energies. As a rule of thumb one can keep in mind that the total energy emitted ($E^2d\phi/dE$), as a function of energy, peaks at about $E=M_\chi/20$ for the intrinsic emission continuum spectrum and about $E=M_\chi/40$ for the cosmological spectrum.
In figure \ref{deboer}, a comparison between cosmological and emission spectra can be seen.

The line signal is different since all photons are emitted at the same energy, $E=M_\chi$ (in the case of a 2$\gamma$-final state) and are observed at the energy $E_0=E(1+z)^{-1}$, depending on at which redshift the WIMPs annihilated. At high redshifts the universe becomes opaque to high energy photons and the signal goes down dramatically at lower $E_0$. This results in the characteristic spectral feature of a sharp cut-off at the WIMP mass, with a tail to lower energies as seen in figure \ref{deboer}.

The left panel of figure \ref{deboer} shows a comparison of the photon spectra at emission and the cosmological spectra from a 60 GeV WIMP only taken into account annihilation into $b\bar{b}$. The crosses are the reanalyzed EGBR of EGRET by \cite{deboer}, where a 60 GeV WIMP has been included in the galactic foreground emission model. Without doing any analysis we note that, in contrasts to claims made in \cite{deboer}, the cosmological spectra, from a 60 GeV WIMP, does not peak at the characteristic 3 GeV bump in the EGBR measurement, where the observed emission spectrum peaks.


\subsection{Halos}

The question of how dark matter is distributed on small, galactic and sub-galactic scale is still a matter of debate. However, N-body simulations show that large structures form by the successive merging of small substructures, with smaller objects usually being denser \cite{diemand}. The density distribution in DM halos, from simulations, are well fitted by simple analytical forms, where the most common one is the NFW profile, \cite{nfw}.
%
%

Since the annihilation rate is proportional to the dark matter density squared, any structure in the DM distribution will significantly boost the annihilation signal from cosmological WIMPs. To take this effect into account we again follow the calculations in \cite{ullio}. The quantity $\Delta^2(z)$ in eq. \ref{flux} describes the averaged squared over-density in halos, as a function of redshift. 

Clumping the DM into halos typically yields a boost of $10^4<\Delta^2(z=0)<10^6$ depending on the choice of halo profile and the model of halo concentration parameter dependence of redshift and halo mass, where we use results from \cite{bullock}. This freedom of choices introduces about a factor of ten each to the uncertainty in the normalization of the cosmological WIMP signal. However, this can be compared to the uncertainty in the signal from point sources, where only the choice of density profile can change the normalization by several orders of magnitude, which is the case with WIMP signals from the galactic center. Another difference, compared to point sources, is that the astrophysics of the halo concentration parameter can change the shape of the $\gamma$-ray spectrum, which is solely determined by particle physics in the case of point sources.

The largest contribution to $\Delta^2(z)$ comes from small halos formed in an earlier, denser universe. However, our understanding of halos at the low mass end is limited due to finite resolution of the N-body simulation. Therefore we have to use a cut-off mass, below which we do not trust our toymodels for the halo concentration parameters. We put this cut-off at $10^5 M_\odot$. Lowering the cut-off might boost the signal even further but will also introduce further uncertainties.



%
\begin{figure}  
\begin{center}
\epsfig{figure=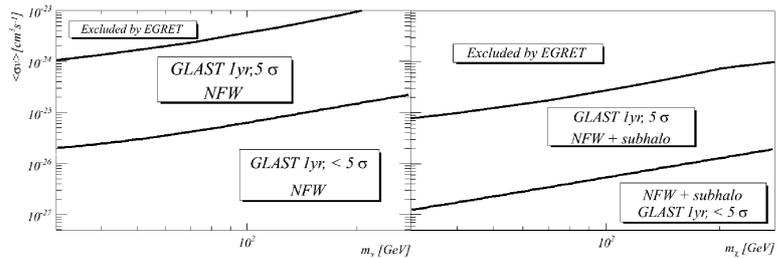,width=12.0cm}
\end{center}
\caption{GLAST 1-year, $5\sigma$ sensitivity for generic, thermal WIMPs annihilation into $b\bar{b}$ and a branching of $10^{-3}$ into two-photon lines. See text for details.} 
   \label{sensitivity1}
\end{figure} 
%

Also within larger halos, N-body simulations indicates that  there should exist smaller, bound halos that have survived tidal stripping. These halos are indicated to have masses all the way down to $10^{-6} \,M_\odot$. 
Although not as massive as the primary  halos the substructure halos arise in higher density environments which makes them denser than their parent halo. The phenomena of halos within halos seems to be a generic feature since detailed simulations reveals substructures even within sub halos \cite{diemand}.

\section{Astrophysical contributions to the EGBR}

\begin{figure}  
\begin{center}
\epsfig{figure=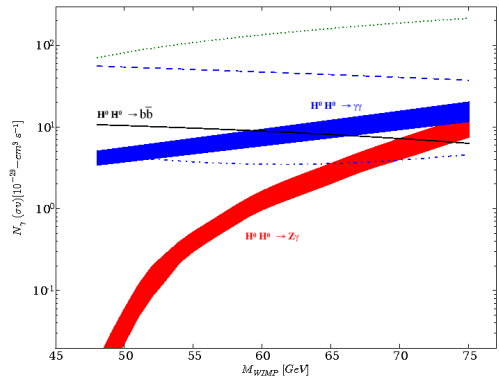,width=6.5cm} %
\end{center}
\caption{The parameter space of IDM shown as photon flux vs. WIMP mass for different annihilation channels \cite{gustafsson}. The dashed line is the 1 year GLAST $5\sigma$ line sensitivity with a NFW profile and the dot dashed line the sensitivity when including substructures.  Points in the parameter space that could be resolved by GLAST are the $\gamma\gamma$ fluxes above the sensitivity lines. The green, dotted line marks the region already excluded by  EGRET, assuming an NFW profile with substructures.}
   \label{sensitivity2}
\end{figure} 
%



The "standard" model for explaining the EGRB is that it consists of diffuse emission from unresolved, $\gamma$-ray point sources such as blazars, quasars, starburst galaxies and starforming galaxies. Contributions from unresolved blazars, consistent with the EGRET blazar catalogue, could account for about 20\% of the measured EGRB at 1 GeV. Taking into account predictions of starburst and starforming galaxies one gets about the measured values of the EGRB at 1 GeV \cite{dermer}. However, these models under-predict the $\gamma$-ray flux at higher energies, arguing for new, hard $\gamma$-ray sources. 

The background used in our sensitivity calculations consists only of unresolved blazars \cite{ullio} where GLAST's increased sensitivity to point sources have been taken into account. In figure \ref{deboer}, this background can be compared to other background models as well as EGRET measurements of the EGBR. Note however that the backgrounds from \cite{dermer} are not treated together in a consistent way, for instance with respect to the optical depth. However, this has been done for the unresolved blazar model and the cosmological WIMPs.

\section{GLAST sensitivity}

Fast detector simulations \cite{scitools} were done for a generic model of WIMPs annihilating into 2$\gamma$ and into $b\bar{b}$ for WIMP masses between 30 and 280 GeV. A $\chi^2$ analysis was performed, assuming that the background consists of unresolved blazars, to obtain a sensitivity plot in $<\sigma v>$ vs $M_\chi$. Also, to the background we added an irreducible contribution from charged particles, at the level of 10 \% of the blazar background. It should be noted that for the calculations presented here we somewhat optimistically assume that we have an ideal extraction of the EGBR as well as a perfect understanding of the conventional astrophysical backgrounds.\\
The WIMP signal was computed using a NFW profile and including the effect of substructures, assuming that they constitute $5\%$ of the mass and have four times the concentration parameter of the parent halo. 
 The result, viewed in figure \ref{sensitivity1}, shows that GLAST is sensitive to total annihilation cross-sections of the order $10^{-24}-10^{-27}$ cm$^3$ s$^{-1}$, depending on the exact halo model, for low masses and about an order of magnitude higher cross section for higher masses.\\
In figure \ref{sensitivity2} the line sensitivity of GLAST to IDM  cosmological WIMPs can be seen. Since the IDM offers so much higher ratio between the line branching and continuum branching the sensitivity was calculated only for the lines as $n_{\gamma,\chi}/\sqrt{n_{bkg}}$. The result is quite dependent on the choice of halo model. For the plain NFW profile GLAST cannot reach a $5\sigma$ level within one year but when adding substructures GLAST is sensitive to almost the entire parameter space of the IDM, at a $5\sigma$ level.  

\section{Conclusion}

We have shown that studying the EGBR with GLAST could offer an interesting way of indirect detection of WIMP DM. Also since the cosmological signal differs in many ways from other point like sources of DM, it will be a useful compliment to such surveys. The level of sensitivity of GLAST still depends on many unknowns, many of them associated with the fact that we do not know enough about the nature of DM. But under our assumptions we find that GLAST will be sensitive to a wide range of interesting WIMP candidates. However, both in the case of the generic and the IDM WIMP the signal needs a light boost for GLAST to be able to cover the most interesting region which could be achieved by adding substructures in the halos. 

\end{document}